\documentclass[prd]{revtex4}
\usepackage{graphicx}

\def\be{\begin{equation}}
\def\ee{\end{equation}}
\def\bea{\begin{eqnarray}}
\def\eea{\end{eqnarray}}

\begin{document}
\title{Electromagnetic Form Factors of the Nucleon \\ 
in Chiral Soliton Models}
\author{Gottfried Holzwarth
}
\affiliation{Fachbereich Physik, Universit\"{a}t Siegen,\\ 
D-57068 Siegen, Germany
}
\begin{abstract}\noindent
The ratio of electric to magnetic proton form factors $G_{\mathrm{E}}^{\mathrm{p}}/G_{\mathrm{M}}^{\mathrm{p}}$
as measured in polarization transfer experiments shows a
characteristic linear decrease with increasing
momentum transfer $Q^2 (< 10$ (GeV/c)$^2$). We present a simple
argument how such a decrease arises naturally in chiral
soliton models. For a detailed comparison of model results with experimentally
determined form factors it is necessary to employ a boost from the soliton rest frame
to the Breit frame. To enforce asymptotic counting rules for form factors,
the model must be supplemented by suitably chosen
interpolating powers $n$ in the boost prescription. 
Within the minimal $\pi$-$\varrho$-$\omega$ soliton model, with the same $n$ for
both, electric and magnetic form factors,
it is possible to obtain a very satisfactory fit to all available proton data
for the magnetic form factor and to the recent polarization results for the ratio 
$G_{\mathrm{E}}^{\mathrm{p}}/G_{\mathrm{M}}^{\mathrm{p}}$.
At the same time the small and very sensitive neutron electric form factor 
is reasonably well reproduced.
The results show a systematic discrepancy with presently available data for the
neutron magnetic form factor $G_{\mathrm{M}}^{\mathrm{n}}$ for $Q^2 > 1$ (GeV/c)$^2$.
We additionally comment on the possibility to extract
information about the form factors in the time-like region and on 
two-photon exchange contributions to unpolarized elastic scattering which
specifically arise in soliton models.
   
\end{abstract} 
\maketitle

\section{Introduction}
Baryons are spatially extended objects.
Soliton models provide spatial profiles for  baryons 
already in leading classical approximation
from the underlying effective action. Therefore all types of
form factors may readily be extracted from soliton models.
Specifically, the wealth of experimental data for electromagnetic 
nucleon form factors pose a severe challenge for chiral soliton models. 

Electron-nucleon scattering experiments which measure ratios of polarization
variables have confirmed that with increasing momentum transfer $Q^2=-q_\mu q^\mu$ 
the proton electric form factor $G_{\mathrm{E}}^{\mathrm{p}}(Q^2)$ decreases 
significantly faster than the proton magnetic form factor $G_{\mathrm{M}}^{\mathrm{p}}(Q^2)$.
This characteristic feature of the electric proton form factor arises naturally
in chiral soliton models of the nucleon and has been predicted previously
from such models~\cite{Ho}. In the following section we give a very simple and
transparent argument for the origin of this result.

We then present a detailed comparison of presently available experimental data with
results from the soliton solution of the minimal $\pi$-$\rho$-$\omega-$meson model.
In Section 1.3 we simply state the relevant classical action for the meson fields without derivation or comment.
It has been discussed extensively in the literature to which we refer. Similarly,
we do not repeat here the derivation of the detailed expressions for the form factors.
We state them explicitly only for the simple purely pionic Skyrme model, and indicate
the modifications brought about by including dynamical vector mesons.

Form factors in soliton models are obtained in the rest frame of the soliton.
A severe source of uncertainty lies in the fact that
comparison with experimental data requires a boost to the Breit frame.
This difficulty applies to all kinds of models for extended objects with internal structure. 
Ambiguities due to differences in boost prescriptions become increasingly significant
for $Q^2$ around and above $(2M)^2$ (with nucleon mass $M$).
In order to enforce superconvergence for $Q^2 \to \infty$, we use in the following a boost prescription with
the same interpolating power $n=2$ for both, electric and magnetic form factors.

In Section 1.5 we then show that within this rather restricted framework it is possible
to obtain a satisfactory fit to the presently available data for the electromagnetic
proton form factors over more than three orders of magnitude of momentum transfer $Q^2$.
This can be achieved with the relevant parameters of the effective action at 
(or close to) their empirical values. 
The electric neutron form factor is a small difference between two larger quantities.
So it is remarkable that the observed $Q^2$-dependence is also essentially reproduced.
The absolute size is closely linked to the effective $\pi$-$\omega$ and $\gamma$-$\omega$
coupling strengths, and it is sensitive to the number of flavors considered.
So it is not difficult to bring also this delicate 
quantity close to the corresponding data. Altogether,
this fit then results in a prediction for the magnetic neutron form factor $G_{\mathrm{M}}^{\mathrm{n}}(Q^2)$.
It turns out that for $Q^2 > 1 $(GeV/c)$^2$ where new data are still lacking, the
calculated result for $G_{\mathrm{M}}^{\mathrm{n}}(Q^2)$ rises above the magnetic proton form factor.
This is in conflict with existing older data.

Prospects to obtain results from soliton models for form factors in the time-like region 
are briefly discussed in Section 1.6. 

Finally, leading contributions to the $2\gamma$-exchange
amplitudes in soliton models are outlined, which may help to reduce
the discrepancies between form factors extracted via Rosenbluth separation
from unpolarized elastic electron-nucleon scattering
and those obtained from ratios of polarization observables.

\section{Characteristic feature of the electric proton form factor}

Chiral soliton models for the nucleon naturally account 
for a characteristic decrease of the ratio $G_{\mathrm{E}}^{\mathrm{p}}/(G_{\mathrm{M}}^{\mathrm{p}}/\mu_p)$ with
increasing $Q^2$. The reason for this behaviour basically originates in the fact 
that in soliton models the isospin for baryons is generated by rotating the soliton in isospace.
The hedgehog structure of the soliton couples the isorotation to a
spatial rotation. Therefore, in the rest frame of the soliton, 
the isovector ($I=1$) form factors
measure the (rotational) inertia density $B_1(r)$, as compared to 
the isoscalar baryon density $B_0(r)$ for the isoscalar ($I=0$) form factors. This
becomes evident from the explicit form of the isoscalar and isovector
form factors in the simple purely pionic soliton model~\cite{Braaten}:
\bea
\label{G0E}
G^{0}_{\mathrm{E}} (k^2) & = & \frac{1}{2}\int d^3r\: j_0(kr) B_0(r)\\
\label{G0M}
G^{0}_{\mathrm{M}} (k^2)/\mu_0 & = & \frac{3}{r_B^2} \int d^3r \: r^2 \; \frac{j_1(k
r)}{kr} B_0(r)\\
\label{G1E}
G^{1}_{\mathrm{E}} (k^2) & = & \frac{1}{2} \int d^3r\: j_0(kr) B_1(r)\\
\label{G1M}
G^{1}_{\mathrm{M}} (k^2)/\mu_1 & = & 3 \int d^3r\: \frac{j_1(kr)}{kr} B_1(r),
\eea
(with mean square isoscalar baryon radius $r_B^2$,
isoscalar and isovector magnetic moments $\mu_0,\mu_1$, 
and normalization $\int B_0(r)d^3r =\int B_1(r)d^3r =1$).

\begin{figure}
\begin{center}
\includegraphics[width=12cm,height=7.5cm,angle=0]{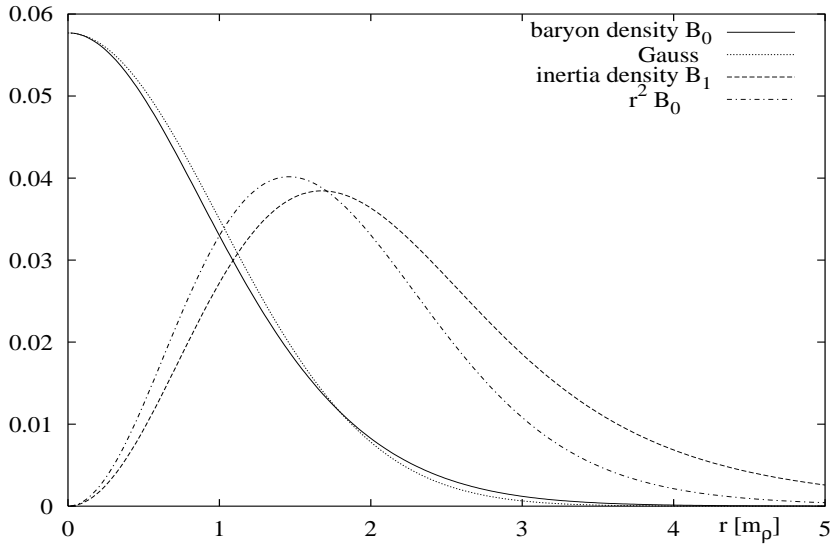}
\end{center}
\caption{Comparison between the topological baryon density $B_0$
and a Gaussian, and between the inertia density $B_1$ and $r^2 B_0$,
for the standard pionic Skyrme model~(\ref{Skyrme})-(\ref{L4}) with $e=4.25$. }
\label{b0b1}
\end{figure}

Evidently, if the inertia density were obtained from
rigid rotation of the baryon density $B_1(r)= (r^2/r_B^2)B_0(r)$,
the normalized isoscalar and isovector magnetic form factors
would satisfy the scaling relation 
\be \label{Mscale}
G^{1}_{\mathrm{M}} (k^2)/\mu_1 = G^{0}_{\mathrm{M}} (k^2)/\mu_0,
\ee
while for the electric form factors the same argument leads to
\be  \label{E1E0}
G^{1}_{\mathrm{E}} (k^2)=-\frac{1}{r_B^2}\left(\frac{\partial}{\partial \vec k}\right)^2\; G^{0}_{\mathrm{E}} (k^2).
\ee
For a Gaussian baryon density $B_0(r)\propto\exp(-(3r^2)/(2r_B^2))$ 
the 'scaling' property (\ref{Mscale}) includes 
also the isoscalar electric form factor 
\be \label{Escale}
G^{1}_{\mathrm{M}} (k^2)/\mu_1 = G^{0}_{\mathrm{M}} (k^2)/\mu_0 =2 G^{0}_{\mathrm{E}} (k^2),
\ee
and Eq.(\ref{E1E0}) then leads to
\be  \label{E1E0Gauss}
G^{1}_{\mathrm{E}} (k^2)=\left(1-\frac{1}{9}k^2r_B^2\right) G^{0}_{\mathrm{E}} (k^2).
\ee
Therefore, for proton form factors
\be \label{protemffs}
G^{\mathrm{p}}_{{\mathrm{E}},{\mathrm{M}}} =G^{0}_{{\mathrm{E}},{\mathrm{M}}} +G^{1}_{{\mathrm{E}},{\mathrm{M}}} , 
\ee
the ratio $G^{\mathrm{p}}_{\mathrm{E}}/(G^{\mathrm{p}}_{\mathrm{M}}/\mu_p)$ resulting from Eqs.(\ref{Mscale}),
(\ref{Escale}) and (\ref{E1E0Gauss}), is 
\be\label{ratio}
R(k^2)=G^{\mathrm{p}}_{\mathrm{E}}(k^2)/(G^{\mathrm{p}}_{\mathrm{M}}(k^2)/\mu_p)=\left(1-\frac{1}{18}k^2r_B^2\right).
\ee
With $r_B^2 \approx 2.3$ (GeV/c)$^{-2}$ $\approx$ (0.3 fm)$^2$, this simple
consideration provides 
an excellent fit (see Fig.\ref{fig1}) through the polarization data for $R(k^2)$.
Of course, in typical soliton models $B_1(r)$ is not exactly proportional 
to $r^2 B_0(r)$ and
the baryon density is not really Gaussian (cf. Fig. \ref{b0b1}). Furthermore,
to compare with experimentally extracted form factors, 
the $k^2$-dependence of the form factors in the soliton rest frame
must be subject to the Lorentz boost from the rest frame to the Breit frame
(which compensates for the fact that typical baryon radii obtained in soliton models are near 0.4-0.5 fm).

But still, we may conclude from these simple considerations
that a strong decrease of the ratio (\ref{ratio})
from  $R=1$ towards an eventual zero near $k^2 \sim 10$ (GeV/c)$^2$
appears as a natural and characteristic feature of 
proton electromagnetic form factors in chiral soliton models.

\section{Chiral $\pi$-$\rho$-$\omega-$meson model}

After the above rather general remarks we consider a specific realistic
model which includes also vector mesons. They are 
known to play an essential role in the coupling of baryons 
to the electromagnetic field and different possibilities
for their explicit inclusion in a chirally invariant
effective meson theory have been suggested~\cite{Kaymak}. 
We adopt the pionic Skyrme model for the chiral SU(2)-field $U$ 
\bea
\label{Skyrme}
{\cal L^{(\pi)}}& = &{\cal L}^{(2)}+{\cal L}^{(4)},\\
\label{L2}
{\cal L}^{(2)}& = &\frac{f_\pi^2}{4}\int \left(-{\mbox tr} L_\mu L^\mu+m_\pi^2 
tr(U+U^\dagger-2) \right)d^3x,\\
\label{L4}
{\cal L}^{(4)}& = &\frac{1}{32e^2}\int {\mbox tr}[L_\mu,L_\nu]^2 d^3x,
\eea
($L_\mu$ denotes the chiral gradients $L_\mu 
= U^\dagger \partial_\mu U$, the pion decay constant is $f_\pi$=93 MeV, and the pion mass $m_\pi$=138 MeV). 
Without explicit vector mesons the
Skyrme parameter $e$ is well established near $e$=4.25. 
Minimal coupling to the photon field is obtained through the local gauge
transformation $U \to \exp(i \epsilon \hat Q)\,U\exp(-i \epsilon \hat Q)$ with the
charge operator $\hat Q=(1/3+\tau_3)/2$. The isoscalar part of the
coupling arises from gauging the standard Wess-Zumino term in the 
SU(3)-extended version of the model.

Vector mesons may be explicitly included as dynamical
gauge bosons. In the minimal version the
axial vector mesons are eliminated in chirally invariant way
\cite{MKW87,SWHH89,Mei93}. This leaves two gauge coupling constants $g_\rho,
g_\omega$ for $\rho$- and $\omega$-mesons,
\be
\label{mini}
{\cal L}={\cal L^{(\pi)}}+{\cal L}^{(\rho)}+{\cal L}^{(\omega)}
\ee
\be
{\cal L}^{(\rho)}= \int \left(-\frac{1}{8} tr \rho_{\mu\nu} \rho^{\mu\nu} 
+\frac{m_\rho^2}{4} tr(\rho_\mu
-\frac{i}{2g_\rho}(l_\mu-r_\mu))^2 \right) d^3x, 
\ee
\be
{\cal L}^{(\omega)}=\int \left(-\frac{1}{4} \omega_{\mu\nu} 
\omega^{\mu\nu} +\frac{m_\omega^2}{2} 
\omega_\mu \omega^\mu +3g_\omega \omega_\mu B^\mu \right) d^3x,
\ee
with the topological baryon current $B_\mu=1/(24 \pi^2)
\epsilon_{\mu\nu\rho\sigma} {\mbox tr} \left(L^\nu L^\rho L^\sigma \right)$, and
$l_\mu=\xi^\dagger \partial_\mu \xi, \;r_\mu=\partial_\mu \xi 
\xi^\dagger$, where $\xi^2=U$. 

The contributions of the vector mesons to the electromagnetic 
currents arise from the local gauge transformations 
\be
\rho^\mu \rightarrow  e^{i \epsilon \hat Q_V}
\rho^\mu e^{-i \epsilon \hat Q_V}  +\frac{\hat Q_V}{g_\rho} 
\partial^\mu \epsilon,~~~~~~~
\omega^\mu \rightarrow  \omega^\mu 
+\frac{\hat Q_0}{g_0} \partial^\mu \epsilon
\ee
(with $\hat Q_0=1/6 \ , \ \hat Q_V=\tau_3 /2$). 
The resulting form factors are expressed in terms of three static and 
three rotationally induced profile functions which characterize the rotating 
$\pi$-$\rho$-$\omega-$hedgehog soliton with baryon number $B=1$. 

Because the Skyrme term ${\cal L}^{(4)}$ at least partly accounts for
static $\rho$-meson effects its strength should be reduced in the presence of
dynamical $\rho$-mesons, as compared to the plain Skyrme model.
The coupling constant $g_\rho$ can be fixed by the KSRF
relation $g_\rho=m_\rho/(2\sqrt{2} f_\pi)$, but small
deviations from this value are tolerable.
The $\omega$-mesons introduce two gauge coupling constants, $g_\omega$
to the baryon current in ${\cal L}^{(\rho)}$, and $g_0$ for the
isoscalar part of the charge operator. 
Within the $SU(2)$ scheme we can
in principle allow $g_0$ to differ from $g_\omega$ and thus 
exploit the freedom in the electromagnetic coupling of the isoscalar
$\omega$-mesons.

The general structure of the form factors as given in Eqs. (\ref{G0E}-\ref{G1M})
for the purely pionic model remains almost unchanged in the $\pi$-$\rho$-$\omega-$model.
In the isoscalar form factors the topological baryon density
$B_0(r)$ is replaced by the total isoscalar charge density. After insertion of the
equation of motion for the $\omega$-mesons we have to replace in Eqs.(\ref{G0E}) and (\ref{G0M})
\be
B_0(r)\Longrightarrow\left(1+\frac{g_\omega}{g_0}(\frac{m_\omega^2}{k^2+m_\omega^2}-1)\right)B_0(r).
\ee
This shows explicitly how the $\omega$-meson pole is introduced into the isoscalar form factors.
For the isovector electric $G^{1}_{\mathrm{E}} (k^2)$ in Eq.(\ref{G1E}) the function $B_1(r)$ again is given by the 
rotational inertia density, which now, however, receives also contributions from the rotationally 
induced $\rho$ and $\omega$
components. In the isovector magnetic $G^{1}_{\mathrm{M}} (k^2)$ in Eq.(\ref{G1M}) the function 
which replaces $B_1(r)$ includes also contributions from the static $\rho$ and $\omega$ profiles
and no longer coincides with the rotational inertia density. 
 The detailed expressions of the form factors which we use here in the minimal $\pi$-$\rho$-$\omega-$model
 (making use of the KSRF relation for $g_\rho$) are given
in Ref.~\cite{Mei93}.  

\section{Boost to the Breit frame}
For all dynamical models of spatially extended clusters
it is difficult to relate the non-relativistic form
factors evaluated in the rest frame of the cluster to the relativistic
$Q^2$-dependence in the Breit frame where the cluster moves with velocity $v$ relative
to the rest frame. For the associated Lorentz-boost factor $\gamma$ we have
\be
\label{Lorentz}
\gamma^2=(1-v^2)^{-1} = 1 + \frac{Q^2}{(2M)^2},
\ee
where $M$ is the rest mass of the cluster. For elastic scattering of clusters composed
of $\nu$ constituents
dimensional scaling arguments  \cite{Matveev} require that the leading power
in the asymptotic behaviour of relativistic form factors is $\sim Q^{2-2\nu}$.
Boost prescriptions of the general form 
\be
\label{boost}
G_{\mathrm{M}}^{\mathrm{Breit}} (Q^2) = \gamma^{-2n_{\mathrm{M}}}\; G^{\mathrm{rest}}_{\mathrm{M}} (k^2),~~~~~~~~~ 
G_{\mathrm{E}}^{\mathrm{Breit}} (Q^2) = \gamma^{-2n_{\mathrm{E}}}\; G^{\mathrm{rest}}_{\mathrm{E}} (k^2)  
\ee
with 
\be\label{map}
k^2=\gamma^{-2}\;Q^2
\ee 
have been suggested with various values for the interpolating
powers $n_{\mathrm{M}},n_{\mathrm{E}}$~\cite{Licht,Ji91}, where $M$ takes the role of an effective mass.

This boost prescription has the appreciated feature 
that a low-$k^2$ region in the rest frame ($0 < k^2 < 1$ (GeV/c)$^2$, say),
where we trust the physical content of the rest frame form factors, appears
as an appreciably extended $Q^2$-regime in the Breit frame. 
So, through the boost (\ref{map}) from rest frame to Breit frame, the region of 
validity of soliton form factors for spatial $Q^2$ is extended.
\begin{figure}[ht]
\begin{center}
\includegraphics[width=9cm,height=4.5cm,angle=0]{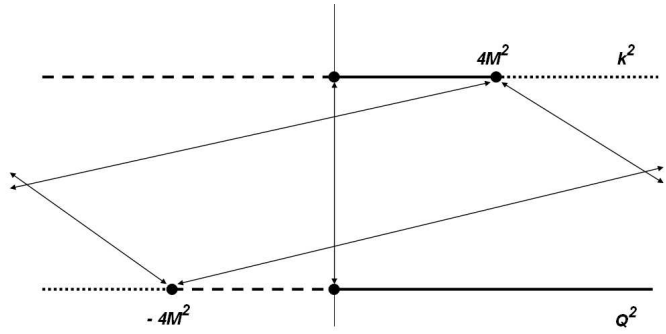}
\end{center}
\caption{The boost (\ref{map})
maps the dashed, solid, and dotted intervals of rest frame $k^2$ onto the dashed, solid, and dotted intervals
of the Breit frame momentum transfer $Q^2$.}
\label{mapfig}
\end{figure}
Evidently, the boost in Eq. (\ref{map})
maps $G^{\mathrm{rest}}(k^2\to 4M^2)~\to~G^{\mathrm{Breit}}(Q^2\to\infty)$.
But, even though $G^{\mathrm{rest}}(4M^2)$ may be very small, it generally
does not vanish exactly. So, unless $n_{\mathrm{M}},n_{\mathrm{E}}\geq 2$, this shows up, of course,
very drastically in the asymptotic behaviour,
if the resulting form factors are divided by the standard dipole
\be
G_{\mathrm{D}}(Q^2)=1/(1+Q^2/0.71)^2,
\ee 
which is the common way to present the nucleon form factors
and accounts for the proper asymptotic $Q^{2-2\nu}$ behaviour of an $\nu=3$ quark
cluster.
So it is vital for a comparison with experimentally determined
form factors for $Q^2 \gg M^2$ to employ a boost prescription
which preserves at least the 'superconvergence' 
property $Q^2G(Q^2)\rightarrow 0$ for $Q^2\rightarrow \infty$.
In accordance with an early suggestion by Mitra and Kumari~\cite{Mitra,Kelly} we use
$n_{\mathrm{M}}=n_{\mathrm{E}} =2$. In any case, the high-$Q^2$
behaviour is not a profound consequence of the model but simply 
reflects the boost prescription. There is no reason
anyway, why low-energy effective models
should provide any profound answer for the high-$Q^2$ limit. Note that the position of 
an eventual zero in $G_{\mathrm{E}}^{\mathrm{Breit}} (Q^2)$ is not affected 
by the choice of the interpolating power $n_{\mathrm{E}}$, and the ratio 
$G_{\mathrm{E}}/G_{\mathrm{M}}$ is independent of the 
interpolating power, as long as $n_{\mathrm{M}}=n_{\mathrm{E}}$.

\begin{figure}[ht]
\begin{center}
\includegraphics[width=13cm,height=9.cm,angle=0]{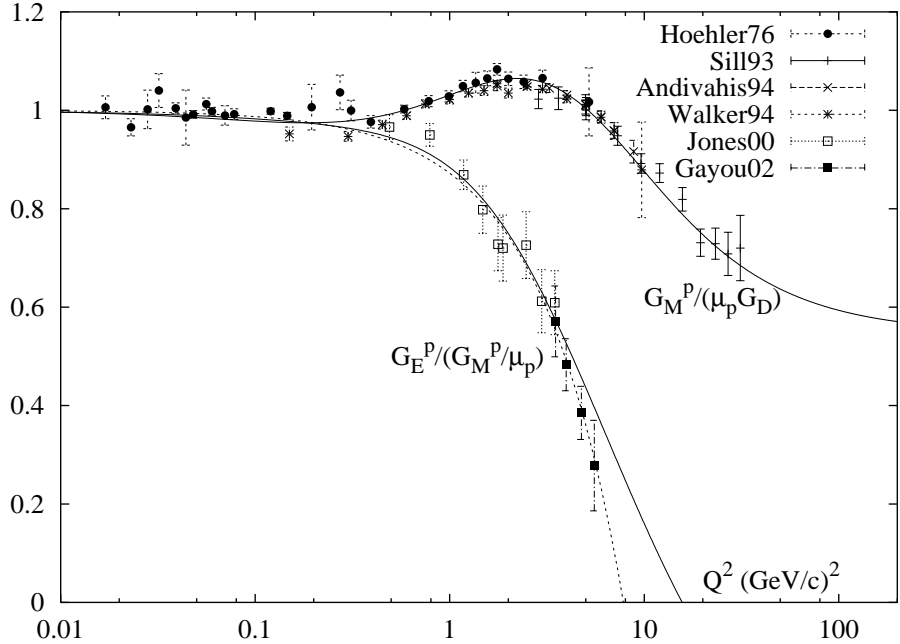}
\end{center}
\caption[]{
Magnetic and electric proton form factors 
$G_{\mathrm{M}}^{\mathrm{p}}/(\mu_p G_{\mathrm{D}})$ 
and $G_{\mathrm{E}}^{\mathrm{p}} /(G_{\mathrm{M}}^{\mathrm{p}}/\mu_p)$ 
for the $\pi$-$\rho$-$\omega-$model 
with the set of parameters given in the text.
The dotted line shows the result of Eq.~(\ref{ratio}) with 
$r_B=0.3$ fm. The abscissa shows $Q^2$(GeV/c)$^2$ on logarithmic scale.
The experimental data are from Refs.~\cite{Hoehler76} -~\cite{Gayou02}.}
\label{fig1}
\end{figure}
\begin{figure}[ht]
\begin{center}
\includegraphics[width=10cm,height=6.5cm,angle=0]{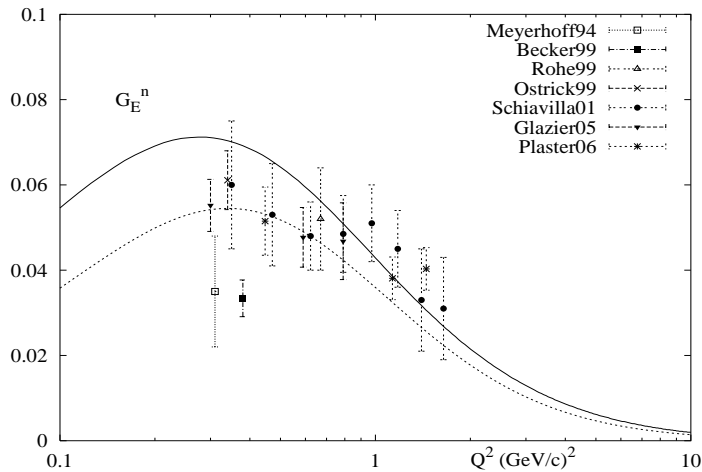}
\end{center}
\caption[]{The neutron electric form factor $G_{\mathrm{E}}^{\mathrm{n}}$
as obtained in the $\pi$-$\rho$-$\omega-$model 
with the set of parameters given in the text.
The dotted line is the standard Galster
parametrization $G_{\mathrm{E}}^{\mathrm{n}} = -\mu_n \tau /(1+5.6 \tau)\cdot G_{\mathrm{D}}$
with $\tau=Q^2/(4M_n^2)$.
Experimental results for $G_{\mathrm{E}}^{\mathrm{n}}$ are mainly from more recent 
polarization data~\cite{Ostrick99} -~\cite{Plaster06}.}
\label{fig3a}
\end{figure}
\begin{figure}
\begin{center}
\includegraphics[width=10cm,height=6.5cm,angle=0]{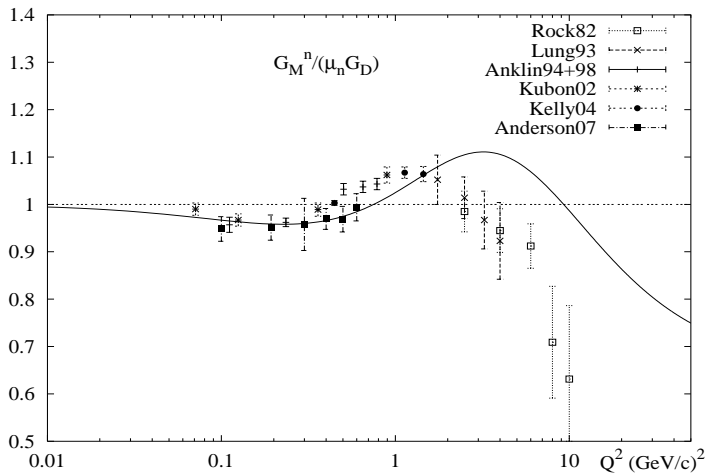}
\end{center}
\caption[]{The magnetic neutron form factor (normalized to the standard dipole)
 $G_{\mathrm{M}}^{\mathrm{n}}/(\mu_n G_{\mathrm{D}})$ in the same model.
Here the data are from Refs.~\cite{Rock} -~\cite{Anderson}.}
\label{fig3b}
\end{figure}

\section{Results}
To demonstrate the amount of agreement with experimental data that 
can be achieved within the framework of such models we present in 
Fig.~\ref{fig1} typical results from the $\pi$-$\rho$-$\omega-$model 
with essential parameters of the model fixed at their physical values:
the pion decay constant $f_\pi=93$ MeV,
the pion mass $m_\pi=138$ MeV,  $\rho$-mass $m_\rho=770$ MeV, $\omega$-mass $m_\omega=783$ MeV,
and the $\pi$-$\rho$-coupling constant at its physical KSRF-value $g_\rho=2.9$.
As variable parameters remain the $\pi$-$\omega$ coupling constant $g_\omega$,
 and the $\omega$-photon coupling constant $g_0$.
Due to the presence of dynamical $\rho$-mesons the strength $1/e^2$ of 
the fourth-order Skyrme term ${\cal L}^{(4)}$ should be reduced as compared to its standard value;
it may even be omitted altogether.
In addition to these three coupling constants, the high-$Q^2$ behaviour of the form factors is,
of course, very sensitive to the effective kinematical mass $M$ which appears in the
Lorentz-boost (\ref{Lorentz}). 

Altogether, while the general features are generic to the soliton model, we use
in the following these four parameters $~g_\omega=1.4,~~g_\omega/g_0=0.75,~~e=7.5$, and $M=1.23$ GeV, 
for the fine-tuning of the proton form factors as shown in Fig.~\ref{fig1}. 
Of course, these four parameters are not independent. 
Changes in the calculated form factors due to variations in one of these parameters 
may be compensated by suitable variations in the others for comparable quality of the fits.
(For example, the agreement shown in Fig.~\ref{fig1} could also be obtained in a three-parameter fit
without Skyrme term (i.e. $1/e^2=0$) with $g_\omega=2.4,~~g_\omega/g_0=0.7$, and $M=1.16$ GeV).

The absolute size of the neutron electric form factor $G_{\mathrm{E}}^{\mathrm{n}}$ 
is closely related to the choice of $g_\omega/g_0$. 
For the chosen set of parameters the maximum of $G_{\mathrm{E}}^{\mathrm{n}}$ exceeds the Galster parametrization
by a factor of about 1.3 (cf. Fig.~(\ref{fig3a})). Correspondingly, the calculated values for
the electric neutron square radii exceed the experimental value by about a factor of 2, and 
we found it difficult to lower them, for reasonable parametrizations within the $SU(2)$ framework. 
But otherwise the shape of $G_{\mathrm{E}}^{\mathrm{n}}$ follows the Galster parametrization rather well, with the 
maximum slightly shifted to lower $Q^2$. 
In the $SU(3)$-embedding of the Skyrme model the mixing coefficients for isoscalar,
isovector, and kaonic contributions to the electromagnetic
form factors cause a sizable reduction of the electric neutron form factor as compared to the $SU(2)$
scheme. The relevant coefficients are listed in Ref.~\cite{Herbert} for the case of exact flavor symmetry;
when symmetry breaking is included, their numerical values reduce the square radius 
$\langle r^2 \rangle ^n_{\mathrm{E}}$ by a factor of about one-half as compared to $SU(2)$,
while the results for the proton remain almost unaffected~\cite{Herbert,Hans}. This cures
the discrepancy for $G_{\mathrm{E}}^{\mathrm{n}}$ in Fig.~(\ref{fig3a}) and for 
$\langle r^2 \rangle ^n_{\mathrm{E}}$ shown in Table 1. However, we are not
aware of calculations of electromagnetic form factors for $Q^2 > 1$(GeV/c)$^2$ 
in the $SU(3)$-embedded Skyrme and vector meson model.

In Fig.\ref{fig3b} we also present the resulting magnetic neutron form factor
$G_{\mathrm{M}}^{\mathrm{n}}$, normalized to the standard dipole $G_{\mathrm{D}}$.
For $Q^2\leq 1$(GeV/c)$^2$ the model result is in perfect agreement with the latest data
\cite{Kelly04}(as quoted in~\cite{Plaster06}),~\cite{Anderson}. For $Q^2 > 1$(GeV/c)$^2$, however, the model prediction
deviates substantially from the available older data~\cite{Rock,Lung}. 
The ratio of the normalized proton and neutron magnetic form
factors $G_{\mathrm{M}}^{\mathrm{n}} \mu_p /(G_{\mathrm{M}}^{\mathrm{p}} \mu_n)$ is independent of the choice of the 
interpolating power $n_{\mathrm{M}}$ in the boost prescription. Therefore
it would be desirable to compare directly with data for this ratio. Experimentally it is 
accessible from quasielastic scattering on deuterium with final state protons and neutrons
detected. The generic scaling relation (\ref{Mscale}) predicts this ratio to be equal to one,
$G_{\mathrm{M}}^{\mathrm{n}} \mu_p /(G_{\mathrm{M}}^{\mathrm{p}} \mu_n) = 1$, so deviations from this 
value indicate, how the function $B_1(r)$ which appears in $G^{1}_{\mathrm{M}} (k^2)/\mu_1$ differs from
$r^2 B_0(r)$ in the specific model considered. Both, the Skyrme model and the $\pi$-$\rho$-$\omega-$model
considered here, consistently predict this ratio to increase above 1 
by up to 15\% for $1 < Q^2$(GeV/c)$^2 < 10$. However, also in this case an $SU(3)$ embedding may change
this prediction appreciably. The presently available
data do not show such an increase for this ratio, in fact they
indicate the opposite tendency. This conflict was already noticed in Refs.~\cite{Ho,Ho02}.
Preliminary data from CLAS~\cite{E94}
apparently are compatible with $G_{\mathrm{M}}^{\mathrm{n}}/(\mu_n G_{\mathrm{D}})=1$ in
the region $1 < Q^2$(GeV/c)$^2 < 4.5$ .

\begin{table}[h]
\caption[]{Nucleon quadratic radii and magnetic moments as obtained
from the chiral $\pi$-$\rho$-$\omega-$model, for the parameters given in the text. 
The experimental values are from Ref.~\cite{Simon}. }
{\begin{tabular}{@{}ccccccc@{}} \toprule
 & $\langle r^2 \rangle ^{\mathrm{p}}_{\mathrm{E}}$ & $\langle r^2 \rangle ^{\mathrm{p}}_{\mathrm{M}}$ & $\langle r^2 \rangle ^n_{\mathrm{E}}$
 & $\langle r^2 \rangle ^n_{\mathrm{M}}$ & $\mu_p$ & $\mu_n$ \\
\colrule
~~~Model~~~& 0.74 & 0.72 & -0.24 & 0.76 & 1.82 & -1.40 \\
~~~Exp.~~~& 0.77 & 0.74 & -0.114 & 0.77 & 2.79 & -1.91 \\
\botrule
\end{tabular}}
\label{tbl1}
\end{table}
In Table~\ref{tbl1} we list quadratic radii and magnetic moments as they
arise from the fit given above. Notoriously low are the
magnetic moments. This fact is common to chiral soliton models and well known.
Quantum corrections will partly be helpful in this respect (see Ref.~\cite{MeiWall97}),
as they certainly are for the absolute value of the
nucleon mass.

Of course, such models can be further extended. Addition of higher-order terms
in the skyrmion lagrangian, explicit inclusion of axial vector mesons,
non-minimal photon-coupling terms,  
provide more flexibility through additional parameters. Our point here,
however, is to demonstrate that a minimal version as described above 
is capable of providing the characteristic features for both
proton form factors and for the electric neutron form factor in remarkable detail. 
In fact, the unexpected decrease of $G_{\mathrm{E}}^{\mathrm{p}}$ was predicted by these models, and it will be
interesting to compare with new data for $G_{\mathrm{M}}^{\mathrm{n}}$ concerning the conflict
indicated in Fig. \ref{fig3b}.

\section{Extension to time-like $Q^2$}
In the soliton rest frame the extension to time-like $k^2$ amounts to finding the 
spectral functions $\Gamma(\nu^2)$ as Laplace transforms of the
relevant densities $B(r)$, e.g. for the isoscalar electric case
\be
r B_0(r)=\frac{1}{\pi^2}\int_{\nu_0^2}^{\infty}e^{-\nu r}\nu\Gamma_0(\nu^2)d\nu,
\ee
and similarly for other cases.
In soliton models the densities are obtained numerically on a spatial grid, therefore
the spectral functions cannot be determined uniquely. Results will always depend on
the choice of constraints which have to be imposed on possible solutions.
But with reasonable choices it seems possible to stabilize the spectral functions
in the regime from the 2- or 3-pion threshold to about two $\rho$-meson masses
and distinguish continuous and discrete structures in this regime~\cite{Ho}.

We note (cf. Fig. (\ref{mapfig})) that the transformation to the Breit frame~(\ref{map}) formally maps
the rest frame form factors $G^{\mathrm{rest}}(k^2)$ for the whole time-like regime $-\infty < k^2 < 0$
onto the Breit-frame form factors $G^{\mathrm{Breit}}(Q^2)$ in the unphysical time-like regime
up to the nucleon-antinucleon threshold $-4M^2 < Q^2 < 0$. On the other hand, the physical
time-like regime $-\infty < Q^2 < -4M^2$ in the Breit frame is obtained as the image of the
spacelike regime $4M^2 < k^2 < \infty$ of form factors in the rest frame. So the 
(real parts) of the Breit-frame form factors
for time-like $Q^2$ beyond the nucleon-antinucleon threshold are formally fixed through Eq. (\ref{boost}). 
However, apart from the probably very limited validity of 
the boost prescription~(\ref{boost}), we do not expect that the form factors
in the soliton rest frame for $k^2 > 4M^2$ contain sufficiently reliable physical information. 
Specifically, oscillations which the rest frame form factors may show for $k^2 \to \infty$, are 
sqeezed by the transformation (\ref{map}) into the vicinity of the physical threshold $Q^2 < -4M^2$.
With $G^{\mathrm{rest}}(k^2)\rightarrow 0$ for $k^2\rightarrow \infty$, the Breit-frame form factors
are undetermined at threshold $Q^2 \rightarrow -4M^2$.

Attempts to obtain form factors for time-like $Q^2$ from soliton-antisoliton configurations
in the baryon number $B=0$ sector face the difficulty that in this sector the only stable
classical configuration is the vacuum. So, any result will reflect the arbitrariness
in the construction of nontrivial configurations. 

Altogether we conclude, that presently we see no reliable way for extracting profound 
information about electromagnetic form factors in the physical time-like regime from soliton models.  

\section{Two-photon amplitudes in soliton models}  

The discrepancies between form factors extracted through the Rosenbluth separation from
unpolarized elastic scattering data~\cite{Qattan} and ratios directly obtained from polarization transfer
measurements~\cite{Jones00,Gayou02} have lead to
the difficult situation that two distinct methods to experimentally determine
fundamental nucleon properties yield inconsistent results~\cite{Arrington}. As a possible remedy,
the theoretical focus has shifted to two-photon amplitudes which enter the
unpolarized cross section and polarization variables in different ways. Two-photon
exchange diagrams involve the full response of the nucleon to doubly virtual
Compton scattering and therefore rely heavily on specific nucleon models. Simple box diagrams
which iterate the single-photon exchange, require virtual intermediate nucleons
and resonances with unknown
off-shell form factors. They have been analysed with various assumptions for the intermediate states
and have been found helpful for a partial reduction of the discrepancies~\cite{2Photbox,Carlson}.

\begin{figure}[h]
\begin{center}
\includegraphics[width =6cm,height=4cm,angle=0]{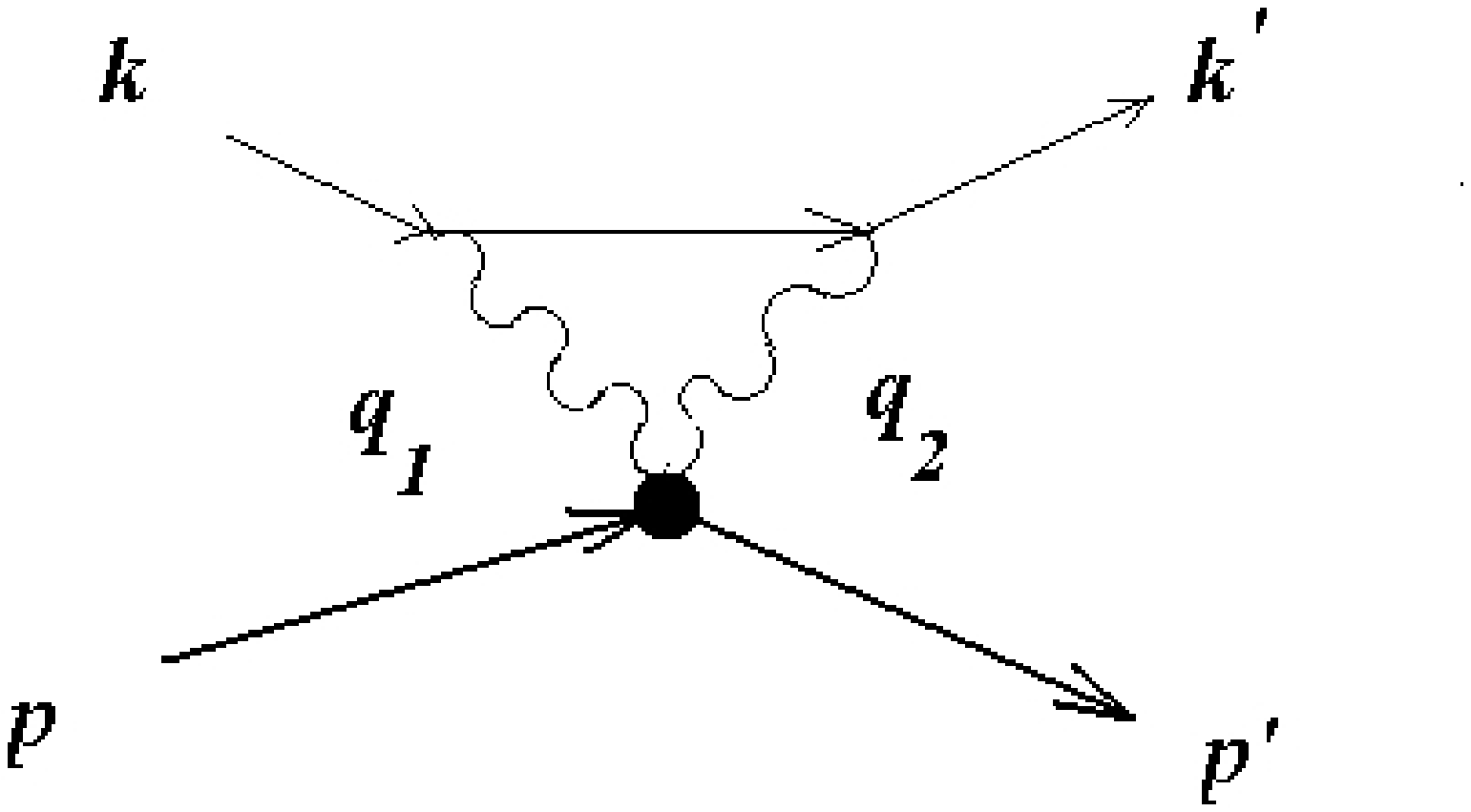}
\end{center}
\caption[]{Electron-nucleon scattering $2\gamma$-exchange amplitude with local
$2\gamma$-soliton vertex with momentum transfer $q=q_1-q_2=k-k'=p'-p.$}
\label{fig4}
\end{figure}

It is interesting to note that, in addition to box diagrams, soliton models contain 
2$\gamma$-exchange contributions where the two virtual photons interact with the pion cloud of the baryon
at {\it local} two-photon vertices.
Products of covariant derivatives
\be
D_\mu U=\partial_\mu U +i[\hat Q,U] A_\mu
\ee
which appear in all terms of the derivative expansion after gauging the chiral fields
with the electric charge $\hat Q$,
naturally produce these local two-photon couplings.
The simplest ones originate  
from the quadratic nonlinear $\sigma$-term and from the gauged Wess-Zumino anomalous action
\be
{\cal L}^{(2\gamma)}_{nl\sigma}=-\frac{f_\pi^2}{4}A_\mu A^\mu 2{\mbox tr}(\hat QU\hat QU^\dagger-\hat Q^2),
\ee
\be
{\cal L}^{(2\gamma)}_{W\!Z}=i\frac{N_c}{48 \pi^2} \varepsilon^{\mu\nu\varrho\sigma}
(\partial_\mu A_\nu) A_\varrho {\mbox tr}\left(\hat Q \partial_\sigma U \hat Q U^\dagger-\hat Q U \hat Q \partial_\sigma U^\dagger
+2\hat Q^2(U^\dagger \partial_\sigma U - U \partial_\sigma U^\dagger)\right).
\ee
After quantization of the collective coordinates the matrix elements of these $2\gamma$-vertices 
sandwiched between incoming and outgoing nucleon states are obtained, without additional parameters, with 
form factors fixed through the soliton profiles. Then the interference terms 
with the single-photon-exchange amplitudes for the unpolarized elastic cross section can be evaluated.
It turns out that the contribution from ${\cal L}^{(2\gamma)}_{nl\sigma}$ interferes only 
with the electric part of the 1-photon-exchange Born term and vanishes after spin averaging. On the other hand,
the scattering amplitude following from ${\cal L}^{(2\gamma)}_{W\!Z}$ interferes only with the magnetic
part of the Born amplitude, so that apart from kinematical factors the unpolarized elastic
electron-nucleon cross section has the general structure
\be
\frac{d\sigma}{d\Omega}\propto \left(
G_{\mathrm{M}}^2(Q^2)+ \frac{\epsilon}{\tau}G_{\mathrm{E}}^2(Q^2)+
\nu (1-\epsilon)G_{\mathrm{M}}(Q^2) F^{(2\gamma)}_{W\!Z}(Q^2) \right)
\ee 
with Lorentz invariants $\tau=Q^2/(4M^2)$, and
\be
\nu=\frac{1}{4}(k+k')\cdot(p+p')= \sqrt{\tau(1+\tau)\frac{1+\epsilon}{1-\epsilon}}~.
\ee 
The form factor $F^{(2\gamma)}_{W\!Z}$ is of the order of the electromagnetic
coupling constant~$\alpha$, 
and involves a loop integral and Fourier transforms of soliton profiles. 
Due to its origin from the Wess-Zumino action, it is parameter free. The possibility to obtain 
parameter free information about the influence of two-photon exchange contributions, makes this 
scheme very attractive. However, it should be mentioned that the infinite part of the
loop integral requires a counterterm which has to be fixed by other experimental input.
This program has been performed in Ref.\cite{Kuhn}. The corrections obtained
have been found to reduce the observed discrepancies, with an absolute size, however, which by itself
is also not sufficient to resolve the problem. It has to be supplemented by iterated 
single-photon exchange.

The $\epsilon$-dependence through $\sqrt{(1+\epsilon)/(1-\epsilon)}$ as contained in $\nu$
is a general symmetry and consistency requirement for the two-photon interaction~\cite{Rekalo}.
There is, however, experimental evidence that within the present error limits the unpolarized 
elastic cross section is consistent with a linear  $\epsilon$-dependence~\cite{Chen,Tomasi}. This still allows
to extract via Rosenbluth separation, effective electric and magnetic form factors which then comprise
also the sum of all relevant 2$\gamma$-contributions. Their ratios may differ appreciably from
ratios of the single-photon-exchange form factors $G^{\mathrm{p}}_{\mathrm{E}}/G^{\mathrm{p}}_{\mathrm{M}}$
as extracted from polarization transfer data, which are believed to remain 
mostly unaffected by  2$\gamma$-contributions~\cite{Carlson}. 
Although at present the situation is not yet fully understood, there is strong evidence that 
2$\gamma$-exchange effects may in fact account for most of the observed differences~\cite{ArMelTj},
and electromagnetic form factors remain the challenging testing ground for models of the nucleon.

The fact that the unexpected results of the polarization transfer experiments follow
as generic consequence from soliton models; that within a minimal specific model form factors 
can be reproduced in detail; and that, in addition to the usual box diagrams,
standard gauging provides a new class of radiative corrections with local 
2$\gamma$-nucleon coupling; all of this once again underlines the strength of the soliton approach to baryons.

\section*{Acknowledgements}
The author is very much indebted to H. Walliser and H. Weigel for numerous discussions.

\end{document}